\documentclass[conference,compsoc]{IEEEtran}

\ifCLASSOPTIONcompsoc
  \usepackage[nocompress]{cite}
\else
  \usepackage{cite}
\fi
\ifCLASSINFOpdf
\else
\fi
\usepackage{amsmath}
\usepackage{amsmath}
\usepackage{amssymb}
\usepackage{mathtools}
\usepackage{bm}
\usepackage{xcolor}
\usepackage{enumitem}
\usepackage{xfrac}
\usepackage{tikz}
\usepackage{graphicx}
\usepackage{caption}
\usepackage{subcaption}
\usepackage{hyperref}
\usepackage{array}
\usepackage{float}
\usepackage{listings}
\usepackage{microtype}

\begin{document}
%


\title{Scaling and Acceleration of Three-dimensional Structure Determination for Single-Particle Imaging Experiments with SpiniFEL}

\author{\IEEEauthorblockN{Hsing-Yin Chang}
\IEEEauthorblockA{
SLAC National Accelerator Laboratory\\
Email: iris@slac.stanford.edu}
\and
\IEEEauthorblockN{Elliott Slaughter}
\IEEEauthorblockA{
SLAC National Accelerator Laboratory\\
Email: eslaught@slac.stanford.edu}
\and
\IEEEauthorblockN{Seema Mirchandaney}
\IEEEauthorblockA{
SLAC National Accelerator Laboratory\\
Email: seema@slac.stanford.edu}
\and
\IEEEauthorblockN{Jeffrey Donatelli}
\IEEEauthorblockA{
Lawrence Berkeley National Laboratory\\
Email: jjdonatelli@lbl.gov}
\and
\IEEEauthorblockN{Chun Hong Yoon}
\IEEEauthorblockA{
SLAC National Accelerator Laboratory\\
Email: yoon82@slac.stanford.edu}
}


%

\maketitle

\begin{abstract}
The Linac Coherent Light Source (LCLS) is an X-ray free electron laser (XFEL) facility enabling the study of the structure and dynamics of single macromolecules. A major upgrade will bring the repetition rate of the X-ray source from 120 to 1 million pulses per second. Exascale high performance computing (HPC) capabilities will be required to process the corresponding data rates. We present SpiniFEL, an application used for structure determination of proteins from single-particle imaging (SPI) experiments. An emerging technique for imaging individual proteins and other large molecular complexes by outrunning radiation damage, SPI breaks free from the need for crystallization (which is difficult for some proteins) and allows for imaging molecular dynamics at near ambient conditions. SpiniFEL is being developed to run on supercomputers in near real-time while an experiment is taking place, so that the feedback about the data can guide the data collection strategy. We describe here how we reformulated the mathematical framework for parallelizable implementation and accelerated the most compute intensive parts of the application. We also describe the use of Pygion, a Python interface for the Legion task-based programming model and compare to our existing MPI+GPU implementation.
\end{abstract}


%
\IEEEpeerreviewmaketitle

\section{Introduction}
\begin{figure}[t]
\centering
\includegraphics[width=0.8\columnwidth]{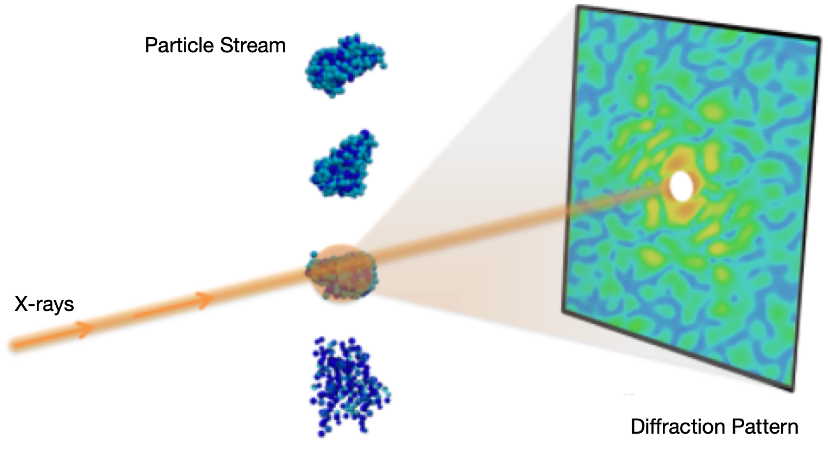}
\caption{SPI involves recording several thousands to millions of single-shot diffraction patterns of identical proteins before its destruction due to ionization and Coulomb explosion. Protein is replenished via a particle stream.}
\label{fig:SPI_exp}
\end{figure}

The high repetition rate and ultrahigh brightness of the Linac Coherent Light Source (LCLS) \cite{LCLS} make it possible to determine the structure of individual molecules, mapping out their natural variation in conformation and flexibility. Structural dynamics and heterogeneities, such as changes in the size and shape of nanoparticles, or conformational flexibility in macromolecules are at the basis of understanding, predicting, and eventually engineering functional properties in the biological, material, and energy sciences. The ability to image the structural dynamics and heterogeneities from noncrystalline proteins is one of the driving forces behind the development of XFEL facilities around the world. Single-particle imaging (SPI) requires femtosecond pulses to outrun radiation damage using a concept called diffraction-before-destruction \cite{chapman2014diffraction}. When an X-ray pulse interacts with a protein freely rotating in space, a diffraction pattern is measured on the detector before the protein is completely destroyed. A train of identical proteins is injected into the interaction region to replenish the sample, as shown in Figure~\ref{fig:SPI_exp}. 

Data analysis must be performed quickly to allow users to guide their data collection strategy by providing feedback on whether sufficient data has been collected and whether the data quality is sufficient to achieve the desired resolution. This will require near-real-time analysis of data bursts (within tens of minutes from the completion of a given burst), requiring commensurate bursts of computational power.

\begin{figure}[t]
\centering
\includegraphics[width=0.8\columnwidth]{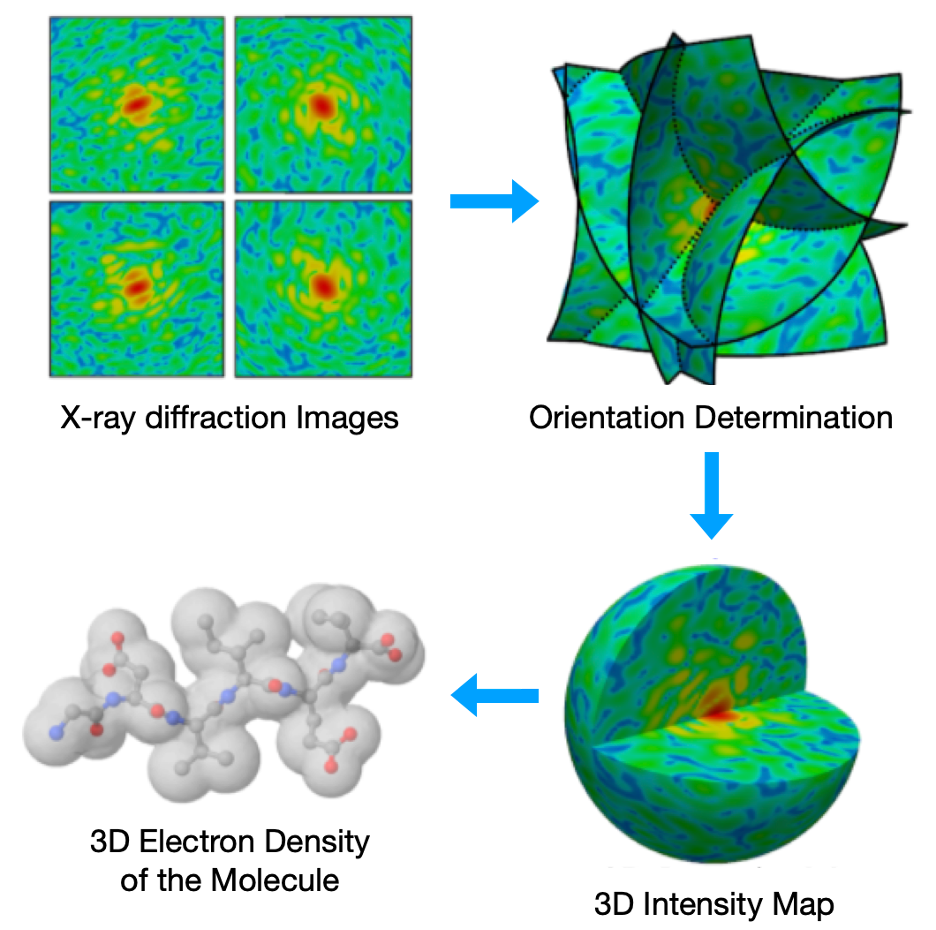}
\caption{Reconstruction requires orientation determination, classification into 2D class averages, and 3D merging of intensities in reciprocal space. The Fourier magnitudes are then used to synthesize a real-space image, but only after recovering the phase information missing from raw data.}
\label{fig:SPI_comp}
\end{figure}

As a part of the U.S. Department of Energy's Exascale Computing Project (ECP), the ExaFEL project \cite{ExaFEL} aims to create an automated analysis pipeline for SPI to transfer diffraction data to supercomputers, where reconstruction is performed, then report results within minutes of data collection. This entails reconstructing a 3D molecular structure from 2D diffraction images by using the multi-tiered iterative phasing (M-TIP) algorithm \cite{donatelli2017reconstruction}. At  a very high-level, each diffraction pattern samples a 2D slice of a protein's 3D Fourier transform. Orientation of the particle has to be determined so that it can be merged in a 3D diffraction volume. The assembled diffraction volume represents the Fourier amplitude of our protein. By solving the phase problem, we recover the 3D electron density of the molecule. The steps are illustrated in Figure~\ref{fig:SPI_comp}.


In Section~\ref{section:framework}, we present algorithmic improvements to the state of the art SPI data analysis. In Section~\ref{section:performance}, we describe GPU offloading and implementation in Pygion, a Python interface to the Legion programming model which we demonstrate on a modest number of nodes with reconstruction results. In Section~\ref{section:results}, we report on computational performance and compare to the MPI+GPU reference implementation.


\section{M-TIP Cartesian-NUFFT Framework}
\label{section:framework}
The existing M-TIP algorithm \cite{donatelli2017reconstruction} relies on a spherical-polar formulation that serially updates the 3D diffraction volume to be consistent with each 2D image, and thus is not scalable. In this section we present a new reformulation of M-TIP based on the nonuniform fast Fourier transform (NUFFT), which allows reconstruction to be parallelized.

A key part of the old algorithm, orientation matching to align 2D images to a 3D model, is based on computing a Wigner-D transform \cite{rossmann1972molecular, kostelec2008ffts} in a spherical basis. Instead, we use a NUFFT of the autocorrelation of the current density model to generate a set of model 2D diffraction images from a fixed orientation grid, and then compare each experimental 2D image to the model image for each orientation.

After alignment, we merge the oriented diffraction patterns into a 3D diffraction volume. Merging is accomplished through the inversion of a NUFFT, which provides an unbiased approach to modelling the experimental data. Furthermore, the availability of fast libraries for computing NUFFTs \cite{greengard2004accelerating, barnett2019parallel}, coupled with iterative linear algebra solvers, provide a scalable approach to recovering the 3D diffraction volume from a large set of 2D diffraction snapshots.

\begin{figure}[H]
\centering
\includegraphics[width=\columnwidth]{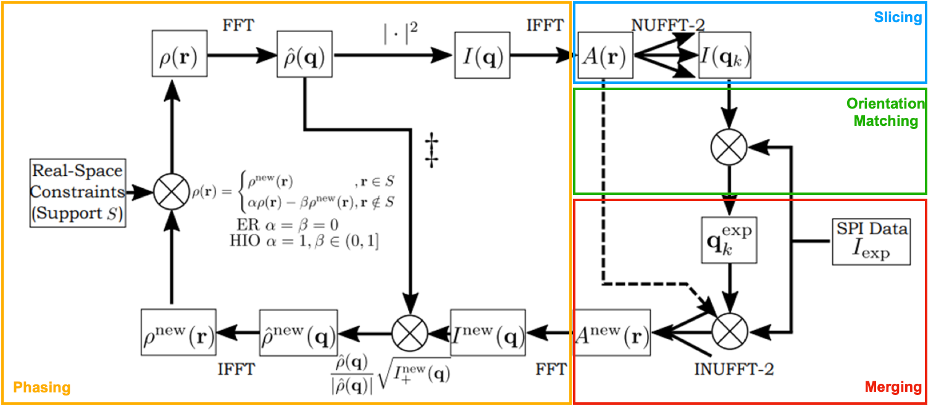}
\caption{Scalable algorithm for SPI. Phasing (yellow) occurs on a single node, but is independent of data size. Slicing (blue), Orientation Matching (green), and Merging (red) scale with number of images. Notation/nomenclature are in Table~\ref{tab:notation}.}
\label{fig:flowchart}
\end{figure}

Scaling the new Cartesian-NUFFT M-TIP framework can be broken down into developing scalable code for four main components:
1) slicing; 
2) orientation matching;
3) merging;
4) phasing, as illustrated in Figure~\ref{fig:flowchart}.
The following subsections describe each of these components.

\begin{table}[H]
\centering
\setlength\extrarowheight{-3pt}
\caption{Notation and nomenclature}
\subcaption*{\scriptsize Quantities}
\scalebox{0.75}{%
\begin{tabular}{lcr}
    $\rho(\textbf{r})$ & Electron Density of Sample \\
    $A(\textbf{r})$ & Autocorrelation of Electron Density \\
    $I(\textbf{q})$ & 3D Intensity Function / Diffraction Volume \\
    $I(\textbf{q}_{k})$ & Model Diffraction Images / Model Orientations \\
    $I_\textbf{exp}$ & Experimental Diffraction Images \\
\end{tabular}}
\medskip
\subcaption*{\scriptsize Grids}
\scalebox{0.75}{%
\begin{tabular}{ll}
    $\textbf{r}$ & $N^{3}$ Cartesian Real-Space Grid ($N = 100 - 500$) \\
    $\textbf{q}$ & $N^{3}$ Cartesian Fourier-Space Grid ($N = 100 - 500$) \\
    $\textbf{q}_{k}$ & Nonuniform Model Data ($10^{5}$ points) \\
\end{tabular}}
\label{tab:notation}
\end{table}

\subsection{Slicing}
In this step, $M$ 2D model images defined on a predetermined set of orientations are computed via the type-2 (uniform to nonuniform, a.k.a. ``forward'') NUFFT of the autocorrelation of the current electron density estimate.

\subsection{Orientation Matching}
Each of the $N$ experimental 2D diffraction images are compared to all of the $M$ model images. The orientation of each experimental image is taken to be that of the model image that is closest in the Euclidean distance metric, which scales as $\mathcal{O}(MN)$. Model images are initialized and distributed at startup to minimize communication.

\subsection{Merging}
Merging can be formulated as inverting a type-2 NUFFT on the oriented experimental data to solve for the autocorrelation of the electron density estimate, which is equivalent to solving a linear, discrete ill-posed problem of the form
\begin{equation}\label{eqn:inverse}
\min_{x} \| Ax-b \|_{2},
\end{equation}
where $A$ is the NUFFT operator, $x$ is a vector representing the autocorrelation on a uniform grid, and $b$ is a vector of intensity data on a nonuniform grid.

However, in SPI experiments, the direct beam necessitates a beam stop to prevent damage to experimental equipment, resulting in a missing region in the detector center that corresponds to low frequency information.
The incompleteness of the data in combination with the high levels of noise lead to instabilities and nonuniqueness of the solution.

The so-called ill-posed problems can be tackled by regularization, where the experimental data are complemented with external or prior information so that the two sources of information together fully determine a unique solution. Note that when splitting up the problem in M-TIP, since we do not have access to all of the priors from constraints on separate tiers, we penalize the perturbation of the model in each tier. Specifically, we use Tikhonov regularization \cite{golub1997tikhonov} to approximate the exact solution, and convert the original linear system into another linear system:
\begin{equation}
A^{*} D A x + \lambda (x-x_{0}) = A^{*}Db,
\end{equation}
where $D$ is weights, $\lambda$ is regularization parameter for the penalty term, and $x_{0}$ the initial guess of the autocorrelation.

The normal equations for the linear system can be expressed as a convolution on a Cartesian grid with dimensions equal to the number of resolution elements per dimension using the Toeplitz method \cite{fessler2005toeplitz}. Setting this up requires two type-1 NUFFTs to be computed, one on the nonuniform experimental data and one with the nonuniform data replaced with 1’s. With this setup, only the type-1 NUFFT computations (computed just once) need to be distributed, since they are a function of the entire experimental dataset. We take advantage of the linearity of the NUFFTs by computing them separately on the experimental data belonging to each node, and use a reduction operation to compute their sum. Once these type-1 NUFFTs are computed, the linear system can then be solved efficiently via conjugate gradient. We repeat this linear solve over nodes, where each rank attempts the solve with a different set of regularization parameters, and use the L-curve method \cite{hansen1999curve} to select the best solution.

\subsection{Phasing}
The phasing step converts a 3D diffraction volume to a molecular structure. Starting from a random \textit{seed} state, we repeatedly apply real-space constraints on $\rho$ and reciprocal-space constraints on $\hat{\rho}$ back and forth through the use of FFT and IFFT. In particular, we apply combinations of error reduction (ER) \cite{gerchberg1972practical}, hybrid input–output (HIO) \cite{fienup1978reconstruction} and shrinkwrap algorithms \cite{marchesini2003x} until convergence is reached.

Phasing only requires computing model values on a Cartesian grid about the same size as the desired density model and can be efficiently solved on just a single node.

\section{Performance Optimization}
\label{section:performance}
In this section, we discuss the acceleration and scaling strategy of our application. 

\subsection{Acceleration -- GPU Offloading}
Merging and slicing operations are offloaded to GPUs using the cuFINUFFT \cite{shih2021cufinufft} library
, and data movement is handled using pyCUDA \cite{kloeckner_pycuda_2012}. We compare performance of a multi-threaded instance using the FINUFFT \cite{barnett2019parallel} library with OpenMP to an equivalent CUDA implementation on a NVIDIA V100 and find that the forward function call runs approximately 1.5$\times$ faster, and the adjoint function call runs approximately 8$\times$ faster for our dataset. A single iteration of merging requires one call to forward, and two calls to adjoint, giving an overall 6$\times$ speedup over a similar CPU-only node.

Orientation matching is accelerated using approximately 800 lines of low-level CUDA code. We found this approach performed better than using existing libraries such as Scikit-learn \cite{scikit-learn} kNN or Facebook AI Similarity Search (FAISS) \cite{JDH17}, by a factor of 50$\times$ and 4$\times$, respectively.

Compared to an equivalent NumPy code \cite{harris2020array}, offloading phasing to GPU using CuPy \cite{nishino2017cupy} gives a 20$\times$ speedup.

\subsection{Parallelization and Scaling -- Pygion}
To distribute the application across multiple CPUs and GPUs on multiple nodes, we use Pygion \cite{slaughter2019pygion}, a Python interface to the Legion task-based runtime system. We chose Pygion because the task-based model is a natural fit for the stages of the computation we described in Section~\ref{section:framework}, and avoids the need for programmers to think about manual data movement or synchronization. Instead, Pygion programs are organized as sets of \emph{tasks} (or functions) that appear to execute sequentially, and \emph{regions}, or data collections passed to tasks. Programmers annotate the \emph{privileges} (read, write, etc.) of the arguments to a task, and the runtime system is responsible for inferring dependencies and issuing data movement as necessary to ensure the program can execute seamlessly on a distributed supercomputer.

Our SpiniFEL implementation in Pygion is structured by turning each major step in the application (as seen in Figure~\ref{fig:flowchart}) into tasks. The data and partitions in our application are straightforward. For example, for the used images, we simply create a region that holds all the images (even if they could not fit on a single node) and then use Pygion's partitioning operators to decompose them into subregions.



\section{Results}
\label{section:results}
We evaluate our approach on Summit \cite{Summit} using a synthetic dataset.
Each Summit node consists of 2 POWER9 CPUs, 6 Nvidia V100 GPUs and 512 GB of memory. Nodes are connected via EDR InfiniBand, and GPUs via NVLink. We consider two implementations, in Legion and MPI, and perform weak scaling experiments up to 32 nodes.

\subsection{Dataset}
\begin{figure}[h!]
\centering
\includegraphics[width=\columnwidth]{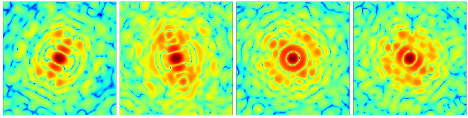}
\caption{Simulated diffraction patterns of a Mm-cpn at random orientations.}
\label{fig:3IYF_data}
\end{figure}

To test our framework, a noise-free synthetic dataset of 500,000 SPI images from a lidless Mm-cpn in the open state from Protein Data Bank (PDB) entry 3IYF is generated, using SPI simulation package skopi \cite{Skopi}. The images provided are 128 $\times$ 128 pixels, with resolution to the edge set to 12 $\AA$.
Figure~\ref{fig:3IYF_data} shows samples of the simulated diffraction patterns.

\begin{figure}[t]
\centering
\includegraphics[width=0.75\columnwidth]{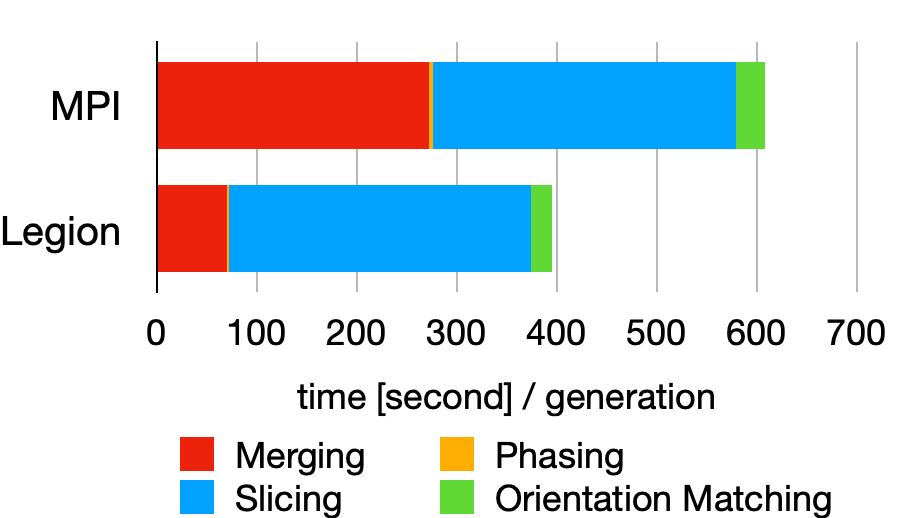}
\caption{Scaling analysis for SpiniFEL proxy app on one node: breakdown of the time spent in each module using MPI and Legion programming model.}
\label{fig:breakdown}
\end{figure}

\begin{figure}[t]
\centering
\includegraphics[width=0.45\columnwidth]{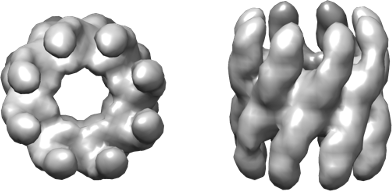}
\caption{Reconstruction of a lidless Mm-cpn in the open state (PDB ID: 3IYF) from simulated noise-free single-particle diffraction data, displayed as density isosurfaces (top and side views) using UCSF Chimera \cite{pettersen2004ucsf}.}

\label{fig:recons}
\end{figure}

\begin{figure}[t!]
\centering
\includegraphics[width=0.7\columnwidth]{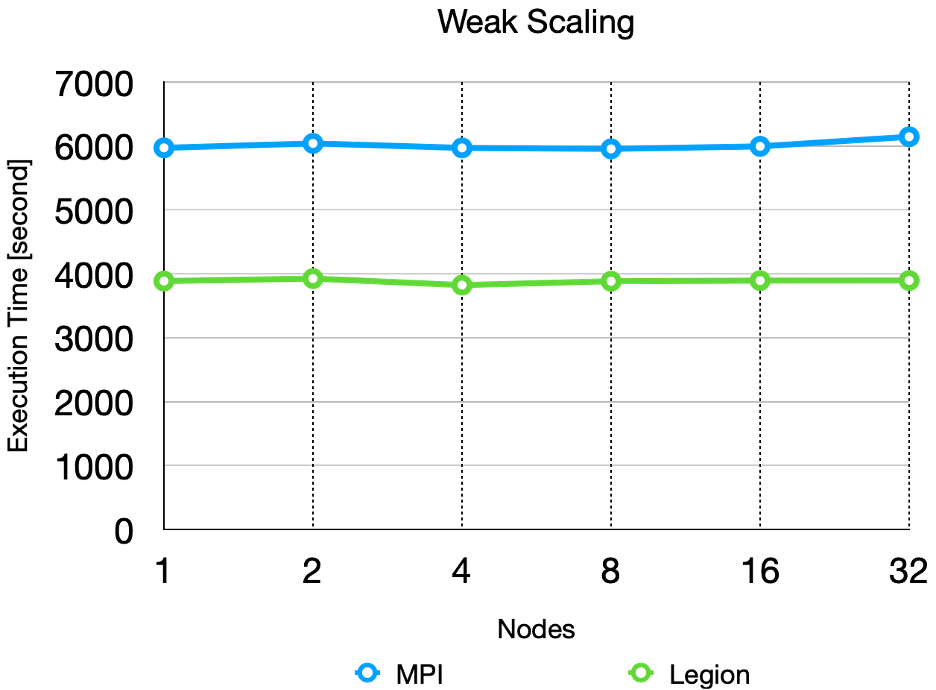}
\caption{Weak scaling in MPI vs. Legion.}
\label{fig:weakscaling}
\end{figure}


\subsection{Performance}
Measurements are performed by distributing a different set of 1,000 simulated diffraction images and the same set of 1,200,000 precomputed model orientations to each node. For weak scaling, we increase the number of nodes from 1 (1,000 images processed) to 32 (for a total of 32,000 images processed) on Summit, with 42 CPUs and 6 GPUs per node.

As shown in Figure~\ref{fig:weakscaling}, both the MPI and Legion versions of SpiniFEL perform well with increasing numbers of nodes as indicated by constant scaling. Compared to an equivalent MPI implementation, Pygion is easier to scale out of the box and leads to speedups of 1.5-fold or more as demonstrated in Figure~\ref{fig:breakdown}, due to its dynamically managed load-balancing of tasks across cores, shared memory (between distinct Python processes on a node) and high level parallelization constructs. The net result is an alternative HPC programming model capable of providing 3D electron density estimates from large scale datasets in about an hour, as shown in Figure~\ref{fig:recons}.



\section{Discussion}
Our MPI and Legion implementations both use an offload model based on a combination of Python (CuPy), C++ (cuFINUFFT) and custom CUDA code. The kernel code, i.e., between MPI send/recv calls or inside Legion tasks, is identical between the two implementations. The main difference is that in MPI these kernels are statically assigned to ranks, whereas in Legion this assignment is controlled by the mapper, a performance tuning interface that exists separately from application code, and therefore is more flexible in many ways, including enabling automatic data layout transformations and task placement decisions.

Our analysis indicates that the performance difference, despite identical code, is due to data layout differences. Legion makes it easier to tune data layout for optimal performance, as this can be done separately from the application code.

\section{Conclusion}
LCLS introduces opportunities for novel science in the domain of SPI. We presented improvements to algorithms in SPI that enable potential scaling to large node counts. These algorithms have been implemented in SpiniFEL, an accelerated and distributed implementation that leverages the Pygion programming model.

\ifCLASSOPTIONcompsoc
  \section*{Acknowledgments}
\else
  \section*{Acknowledgment}
\fi

This research was supported by the Exascale Computing Project
(17-SC-20-SC), a collaborative effort of the U.S. Department
of Energy Office of Science and the National Nuclear Security
Administration. Use of the Linac Coherent Light Source (LCLS),
SLAC National Accelerator Laboratory, is supported by the U.S.
Department of Energy, Office of Science, Office of Basic Energy
Sciences under Contract No. DE-AC02-76SF00515.



%

\bibliographystyle{unsrt}
\bibliography{references}

\end{document}